\documentclass[12pt]{article}

\usepackage{graphicx} \usepackage{natbib} 

\title{The value of human and machine in machine-generated creative contents}
\author{Weina Jin}
\date{}

\begin{document}

\title{The value of human and machine in machine-generated creative contents}

\author{Weina Jin}

\date{}

\maketitle
\begin{abstract}
The seemingly “imagination” and “creativity” from machine-generated contents should not be misattributed to the accomplishment of machine. They are accomplishments of both human and machine. Without human interpretation, the machine-generated contents remain in the imaginary space of the large language models, and cannot automatically establish grounding in the reality and human experience. \end{abstract}

Does machine have creativity? Can AI imagine? What is the human value in the era of generative artificial intelligence (AI)? These are the questions people may ponder upon when seeing the fictional images or texts that large language models (LLMs) generate, such as chatGPT or deepseek. These questions are significant because, if machines can “create” or “imagine,” it will raise doubt about what creativity, imagination, and art really are, and what’s the point of human performing creativity, imagination, and art if machines can also do them. 

\begin{figure}[ht]
    \centering
    \includegraphics[width=0.8\linewidth]{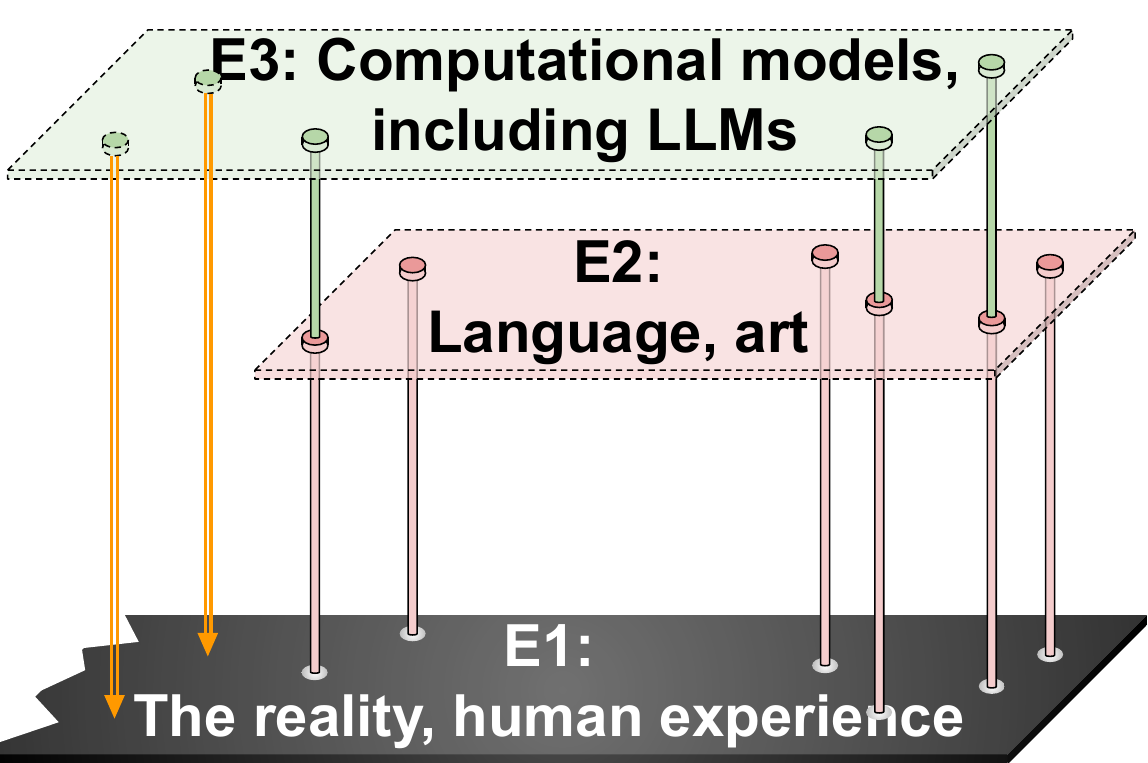}
    \caption{
Visualization of the relationship of the three entities \textbf{E1. reality and human experience}, \textbf{E2. language and art}, and \textbf{E3. LLMs}. E1 is visualized as the ground. E2 and E3 are visualized as the floors built upon the ground as abstract representations. The entities do not completely overlap with each other, which visualizes \textbf{P1 irreducibility} that a representation cannot replace the original complex entity. The dots with solid outlines and their corresponding pillars supporting a floor represent the grounding instances that make the abstract representation valid (\textbf{P2. valid representation}). A valid representation space is the imaginary space (i.e., floors with dashed outlines) spanned by the solid dots. The dots with dashed outlines indicate newly inferred instances that fulfill \textbf{P3. symbolic authenticity}. The dashed, not solid, outlines indicate that the validly inferred new instances within the representation space do not automatically ground in a lower-level phenomenon, unless an explicit grounding connection is made (visualized as the vertical arrow).
}
\label{fig:grounding}
\end{figure}

In this article, I will inspect the above questions by analyzing and understanding the capability and limits of machine. My analysis comes from ontological and epistemological perspectives that concern what we know about the real world and how we know it~\citep{CrasnowSharonL2024Feap}. LLMs are a type of AI models that recognize complex patterns from large-scale training data by training to be a next word or next image patch predictor. Like any computational models, LLMs are abstract representations of the phenomenon they represent. LLMs are trained on human language or image data to mimic the underlying phenomenon —- language or visual art —- represented in the training data. The human language or visual art can also be regarded as an abstract entity that represents the phenomenon of reality and human experience. The three entities \textbf{E1. reality and human experience}, \textbf{E2. language and art}, and \textbf{E3. LLMs} can be regarded as forming a representation spiral, with each latter entity is a mimicking and abstract representation of its former entity. The relationship can be visualized in Fig.1. There are several properties (P1-P4) about the abstract representation: 

\textbf{P1. [Irreducibility]} \textit{A complex phenomenon is irreducible and cannot be fully represented.}

P1 indicates that if an entity is a complex phenomenon, it is incompressible and cannot be reduced to a representation without losing information. Any of its representation is always incomplete and cannot replace the original phenomenon~\citep{10.1093/oso/9780198821939.003.0001,CilliersPaul2016Cc}. 

\textbf{P2. [Valid representation]} \textit{An abstract space is a valid representation of the target phenomenon, if the space is spanned by instances that have established grounding with the target phenomenon.}

For \textbf{E2. language and art} to be a valid representation of \textbf{E1. the reality and human experience}, they should consist of instances of creative writings or art works that are created by humans grounding in our living experiences in the real world. For \textbf{E3. LLMs} to be a valid representation of \textbf{E2. language and art}, the LLMs should pass tests to show the LLMs have learned the training data distribution from the \textbf{E2. language and art} entity of human languages and art works. 

\textbf{P3. [Symbolic authenticity]} \textit{For a representation, it can perform inference within its abstract representation space, as long as the inference follows rules and assumptions that make the abstract representation hold at the first place. We call such inferences fulfilling symbolic authenticity.}

“Symbolic authenticity” uses the word “symbol” because the inference is performed in the abstract symbolic space. “Authenticity” denotes that the inference is valid. P3 indicates that the abstract representation space itself can be regarded as an imaginary space to generate new thoughts or instances by validly inferring within the space. 

\textbf{P4. [Grounding authenticity]} \textit{For newly-inferred instances within an abstract representation space that fulfill symbolic authenticity, they do not automatically establish grounding with the phenomenon it represents. We call that symbolic authenticity doesn’t equal grounding authenticity. For such instances to have grounding authenticity, they should establish grounding in the target phenomenon.}

The reason that “symbolic authenticity doesn’t equal grounding authenticity” lies in P1. Irreducibility. Because the target phenomenon cannot be fully described and replaced by its abstract representation, the imaginary instances in the representation space may not have correspondence (i.e., grounding) in the phenomenon. For example, the phrase “putting theory into practice” is a common sense because theory is an abstract representation of the reality. What works in theory (i.e., symbolic authenticity) may not necessarily work in practice (i.e., grounding authenticity).

\vspace{1.5em}

We can use the above perspective to understand LLMs. First, regarding the value of the machine, because both LLMs and language/art are imaginary entities, and LLMs mimic language/art, LLMs that establish valid representation of language/art can be regarded as an extension of the language/art space. This aspect shows the value and benefits of LLMs: LLMs enable us to explore the vast imaginary space of language/art or generate new imagery language/art spaces. By facilitating the expansion of our imaginations, this value of LLMs strengthens the original value and benefit of art and creative writing that enables us to explore the imaginary space. 

Second, despite the benefit of expanding the imaginary space, LLMs have intrinsic limits that language/art doesn’t have. The new instances in the language/art space can easily establish grounding in the reality and human experience, because the new instances are created by humans and are interpreted by humans to relate the new instances to certain human experiences. This obvious fact, however, doesn’t hold for LLMs. For newly-generated instances from LLMs, they are not inherently  grounded in the reality and human experience due to the above properties P1 and P4: the abstract representation of LLMs cannot be a replacement of the underlying phenomena they mimic (i.e., E1 and E2). Thus, the new instances in the LLMs space don’t automatically establish correspondences in the reality and human experience, and lack their roots in the living experience in the real world. We call it the groundlessness limit of LLMs. 

The groundlessness limit has broad implications for machine-generated contents. Because these contents cannot automatically ground in the reality and living experiences, we cannot regard the contents alone as “art” that shows “creativity” or “imagination” of the machine. Rather, as we analyzed in the above first point, the “imaginations” exhibited in the machine-generated contents are explorations in the imaginary space of the LLM models that can help to extend our human imagination in the forms of texts and images. Creativity, imagination, language, and art have richer meanings than merely exploring the imaginary spaces in the LLMs models. At least some of their meanings include being able to freely translate in-between the imaginary space and the real world: such as relating a phenomenon in the real world to something imaginary, and drawing inferences from the imaginary spaces to the concrete living experiences. The reason that we may leave with the impression of regarding LLM-generated texts/images as the manifestation of the “imagination” or “creativity” of machine is because we humans implicitly close the loop for LLMs from real-world phenomena to the LLM model space and back to the real world, such that the machine-generated contents appear to exhibit “imagination” or “creativity” as art works or creative writings. The invisible, often neglected human work to “close the loop” for LLMs include: 

1. Humans create, select, and label creative contents of texts and images that are used as training data for LLMs. Humans provide preference for machine-generated contents as feedbacks in the training of LLMs (i.e., reinforcement learning from human feedback), and embed implicit knowledge about the phenomena and training data in the model design and training process. These knowledge forms the epistemic basis for LLMs. As information agents, LLMs depend on these human works to represent valid representations about the training data~\citep{jin2025aiimagination}. 

2. When generating a machine-generated content, humans provide the prompts that are based on real-world phenomena and/or human experience. When viewing machine-generated contents, humans implicitly draw connections of the contents to the real world and our living experience. This human interpretation process is the key to establish grounding of the machine-generated contents in reality and human experience. With human interpretation, the machine-generated contents fulfill grounding authenticity and encode rich meaning to us. In other words, LLMs as disembodied abstract computational models do not generate meanings. It is the human interpretation process that generates meaning (i.e., the grounding in reality and human experience) for the machine-created contents.

My analysis shows that attributing “imagination,” “art,” or “creativity” to LLMs is an overclaim and an anthropomorphic tendency~\citep{ibrahim2025thinkinganthropomorphicparadigmbenefits,10.5555/3692070.3692121}, which should be avoided. LLMs as symbolic representations do not inherently ground in the reality and living experience as the words “imagination,” “art,” or “creativity” encode. The value of LLMs lies in their ability to expand our imaginary spaces of texts and visuals encoded in their vast collections of training data. The value of humans lies in our implicit grounding of the machine-generated contents in our living experiences. Arts, imaginations, and creativity enable us to experience the living world in rich and meaningful ways. Understanding the respective value of human and machine can help us use LLMs to enhance, rather than impoverish, our living experience with arts and creative writing. 

\bibliography{ai_art.bib} 

\bibliographystyle{plainnat}

\end{document}